\documentclass[conference]{IEEETran}
\usepackage{amsmath,amssymb,amsthm}
\usepackage{newalg}
\usepackage{multirow}
\usepackage{multicol}
\usepackage{graphicx}
\usepackage{stmaryrd}
\usepackage{blkarray}
\usepackage{subcaption}

\DeclareMathOperator{\F}{\mathbb F}
\DeclareMathOperator{\W}{\mathbf W}
\DeclareMathOperator{\bS}{\mathbf S}

\theoremstyle{plain}

\newtheorem{remark}{Remark}

\newcommand{\mF}{\mathcal F}
\newcommand{\set}[1]{\left\{{#1}\right\}}

\DeclareMathOperator{\sgn}{sgn}

\begin{document}

\title{Efficient decoding of polar codes with some 16$\times$16 kernels}
\author{
\IEEEauthorblockN{Grigorii Trofimiuk, Peter Trifonov}
\IEEEauthorblockA{Saint Petersburg Polytechnic University\\
Email: \{grigoriyt,petert\}@dcn.icc.spbstu.ru}}

\maketitle

\begin{abstract}
A decoding algorithm for polar codes with binary $16\times 16$  kernels with polarization rate $0.51828$ and scaling exponents $3.346$ and $3.450$ is presented. The proposed approach exploits the relationship of the considered kernels and the Arikan matrix to significantly reduce the decoding complexity  without any performance loss. Simulation results  show that polar (sub)codes with $16\times 16$ kernels can outperform polar codes with Arikan kernel, while having lower decoding complexity. 
\end{abstract}

\section{Introduction}

Polar codes are a novel class of error-correcting codes, which  achieve the symmetric capacity of a binary-input discrete memoryless channel  $W$, have low complexity construction, encoding and decoding algorithms \cite{arikan2009channel}. 
However, the performance of polar codes of practical length is quite poor. The reasons for this are the presence of imperfectly polarized subchannels and the suboptimality of the successive cancellation (SC) decoding algorithm. To improve performance, successive cancellation list decoding (SCL) algorithm \cite{tal2015list}, as well as various code constructions were proposed
 \cite{trifonov2016polar,trifonov2017randomized,wang2016paritycheckconcatenated}.  

Polarization is a general phenomenon, and is not restricted to the case of Arikan matrix \cite{korada2010polar}.  One can replace it by a larger matrix, called \textit{polarization kernel}, which can provide higher polarization rate. Polar codes with large kernels were shown to provide asymptotically optimal scaling exponent \cite{fazeli2018binary}. Many kernels with various properties were proposed \cite{korada2010polar,fazeli2014scaling,presman2015binary,buzaglo2017efficient}, but, to the best of our knowledge, no efficient decoding algorithms for kernels with polarization rate greater than $0.5$ were presented, except \cite{miloslavskaya2014sequentialBCH}, where an approximate algorithm was introduced. Therefore, polar codes with large kernels are believed to be impractical due to very high decoding complexity.

In this paper we present  reduced complexity decoding algorithms for $16\times 16$ polarization kernels with polarization rate  $0.51828$ and scaling exponents $3.346$ and $3.45$. We show that with these kernels increasing list size in the SCL decoder provides much more significant performance gain compared to the case of Arikan kernel, and  ultimately the proposed approach results in lower decoding complexity compared to the case of polar codes with Arikan kernel with the same performance.

The proposed approach exploits the relationship between the considered kernels and the Arikan matrix. Essentially, the log-likelihood ratios (LLRs) for the input symbols of the considered kernels are obtained from the LLRs computed via the Arikan recursive expressions. 

\section{Background}
\label{sBackground}
\subsection{Channel polarization}
Consider a binary input memoryless channel  with transition probabilities $W\{y|c\}, c\in \F_2, y\in \mathcal Y$, where $ \mathcal Y$ is output alphabet. For a positive integer $n$, denote by $[n]$ the set of $n$ integers $\{0,1,\dots\,n-1\}$. A \textit{polarization kernel} $K$ is a binary invertible $l \times l$ matrix, which is not upper-triangular under any column permutation. The Arikan kernel is given by $F_2=
\begin{pmatrix}
1&0\\
1&1
\end{pmatrix}. $
%
 An $(n = l^m, k)$ polar code is a linear block code generated by $k$ rows of matrix $G_m = M^{(m)}K^{\otimes m}$, where  $M^{(m)}$ is a digit-reversal permutation matrix, corresponding to mapping  $ \sum_{i = 0}^{m-1}t_il^i \rightarrow  \sum_{i = 0}^{m-1}t_{m-1-i}l^i$,$t_i \in [l]$.
The encoding  scheme is given by
$ c_0^{n-1}= u_0^{n-1}G_m$,
where $u_i,i\in \mathcal F$ are  set to some pre-defined values, e.g. zero (frozen symbols),  $|\mF| = n - k$, and the remaining values $u_i$ are set to the payload data. 

It is possible to show that a binary input memoryless channel $W$ together with matrix $G_m$ gives rise to bit subchannels $W_{m,K}^{(i)}(y_0^{n-1},u_0^{i-1}|u_i)$ with capacities approaching $0$ or $1$, and fraction of noiseless subchannels approaching $I(W)$ \cite{korada2010polar}. Selecting  $\mF$ as the set of indices of low-capacity subchannels enables almost error-free communication. It is convenient to define probabilities \begin{align}
 \label{mKernelStepW}
 W^{(i)}_{m,K}(u_0^{i}|y_0^{n-1})=&\frac{W_{m,K}^{(i)}(y_0^{n-1},u_0^{i-1}|u_i)}{2W(y_0^{n-1})}\nonumber\\
  = &\sum_{u_{i+1}^{n-1}}\prod_{i = 0}^{n-1}W((u_0^{n-1}G_m)_i|y_i).
 \end{align}
Let us further define  $\W^{(j)}_{m}(u_0^{j}|y_0^{n-1}) = W^{(j)}_{m,K}(u_0^{j}|y_0^{n-1})$, where  kernel $K$ will be clear from the context. We  also need  probabilities  $W^{(j)}_{t}(u_0^{j}|y_0^{l-1}) = W^{(j)}_{1,F_2^{\otimes t}}(u_0^{j}|y_0^{l-1})$ for Arikan matrix  $F_2^{\otimes t}$. Due to the recursive structure of $G_n$,
one has
\begin{align}
\W^{(sl+t)}_{m}(u_0^{sl+t}|y_0^{n-1}) = 
\quad \quad \quad \quad \quad \quad \quad \quad \quad \quad \quad \quad \quad \nonumber\\ 
\sum_{u_{sl+t+1}^{l(s+1)-1}} \prod_{j = 0}^{l-1} \W_{m-1}^{(s)}
(\theta_K[u_0^{l(s+1)-1},j]|y_{j\frac{n}{l}}^{(j+1)\frac{n}{l}-1})
\end{align}
where $\theta_K[u_0^{(s+1)l-1},j]_r = (u_{lr}^{l(r+1)-1}G_n)_j, r \in [s+1]$. A trellis-based algorithm for computing these values was presented in \cite{griesser2002aposteriori}.

 At the receiver side, one can successively estimate 
 \begin{equation}
 \label{mSCProb}
 \widehat u_i=\begin{cases}\arg\max_{u_i\in \F_2} \W_m^{(i)}(\widehat u_0^{i-1}.u_i|y_0^{n-1}), &i\notin\mF,\\
\text{the frozen value of $u_i$}&i\in \mF.
\end{cases}
\end{equation}
 This is known as the successive cancellation  (SC) decoding algorithm. 

\section{Computing kernel input symbols LLRs}
\subsection{General case}
Our goal is to compute efficiently probabilities $\W^{(i)}_{m}(u_0^{i}|y_0^{n-1})$ for a given polarization transform $K^{\otimes m}$. Let us assume for the sake of simplicity that $m=1$.
The corresponding task will be referred to as {\em kernel processing}.

 We propose to introduce approximate  probabilities 
\begin{align}
\widetilde \W_1^{(j)}(u_0^j|y_0^{l-1}) &= 
\max_{u_{j+1}^{l-1}} \W^{(l-1)}_{1}(u_0^{l-1}|y_0^{l-1}) \quad \quad \quad \nonumber \\
&=\max_{u_{j+1}^{l-1}}\prod_{i = 0}^{l-1}W((u_0^{l-1}K)_i|y_{i}).
\label{mKernelWApprox}
\end{align}
%
This is the probability of the most likely continuation of path $u_0^j$ in the code tree, without taking into account possible freezing constraints on symbols $u_i,i>j$. Note that the same probabilities were introduced in \cite{miloslavskaya2014sequentialBCH,miloslavskaya2014sequential}, and shown to provide substantial reduction of the complexity of sequential decoding of polar codes. 

Decoding can be implemented using the log-likelihood
ratios
$\bar \bS_{m,i}=\bar \bS_m^{(i)}(u_0^{i-1}|y_0^{n-1})=
\ln\frac{\W_m^{(i)}(u_0^{i-1}.0|y_0^{n-1})}{\W_m^{(i)}(u_0^{i-1}.1|y_0^{n-1})}.$ Hence,  kernel output LLRs $\bar \bS _{1,i}, i \in [l]$ can be approximated by  
\begin{align}
\label{mKernelLog}
\bar \bS_{1,i} \approx \bS_{1, i}   = \ln \frac{\widetilde \W_1^{(i)}(u_0^{i-1}. 0|y_0^{l-1})}{\widetilde \W_1^{(i)}(u_0^{i-1}. 1|y_0^{l-1})} \quad \quad \quad \quad \quad \quad \quad \quad \quad \quad
\nonumber\\
=\max_{u_{i+1}^{l-1}}\ln \W^{(l-1)}_{1}(u(0)^i|y_0^{l-1})
 - \max_{u_{i+1}^{l-1}}\ln \W^{(l-1)}_{1}(u(1)^i|y_0^{l-1}), 
\end{align}
where $b(a)^{i} = (b_0^{i-1}.a.b_{i+1}^{l-1})$. 
The above expression means that $\bS_{1, i} $ can be computed by performing ML decoding of the code, generated by last $l-i+1$ rows of the kernel $K$, assuming that all $u_j,i<j<l,$ are equiprobable.

%

\subsection{Binary algorithm}
\label{sLLRSimple}
Straightforward evaluation of \eqref{mKernelLog} for arbitrary kernel has complexity $O(2^ll)$. However, we have a simple explicit recursive procedure for computing these values for the case of the Arikan matrix $F_2^{\otimes t}$.

Let $l = 2^t$. Consider encoding scheme $c_0^{l-1} = v_0^{l-1}F_2^{\otimes t}$. Similarly to \eqref{mKernelWApprox}, define approximate probabilities $$\widetilde W_t^{(i)}(v_0^i|y_0^{l-1})=
\max_{v_{i+1}^{l-1}} W_t^{(l-1)}(v_0^{l-1}|y_0^{l-1})
$$
and modified log-likelihood ratios $$S_\lambda^{(i)}(v_0^{i-1},y_0^{l-1})=\log\frac{\widetilde W_\lambda^{(i)}(v_0^{i-1}.0|y_0^{l-1})}{\widetilde W_\lambda^{(i)}(v_0^{i-1}.1|y_0^{l-1})}.$$

It can be seen that 
\begin{align}
S_{\lambda}^{(2i)}(v_0^{2i-1},y_0^{N-1})
=&Q(a,b)\label{mMinSum1}\\
S_{\lambda}^{(2i+1)}(v_0^{2i},y_0^{N-1})=&P(a,b,v_{2i})
,\label{mMinSum2}
\end{align}
where $N=2^{\lambda}$, $a=S_{\lambda-1}^{(i)}(v_{0,e}^{2i-1}\oplus v_{0,o}^{2i-1},y_{0,e}^{{N}-1})$, $b=S_{\lambda-1}^{(i)}(v_{0,o}^{2i-1},y_{0,o}^{N-1})$, $Q(a,b) =\sgn (a)\sgn (b)\min(|a|,|b|)$, $P(a,b,c) = (-1)^{c}a+b$. Then the log-likelihood of a path (path score) $v_{0}^i$ can be obtained as  \cite{trifonov2018score}
\begin{align}
R(v_0^i|y_0^{l-1})=&\log\widetilde W_t^{(i)}(v_0^i|y_0^{l-1})\nonumber\\
=R&(v_0^{i-1}|y_0^{l-1})+\tau\left(S_t^{(i)}(v_0^{i-1},y_0^{l-1}),v_i\right),\label{mScoreMinSum}
\end{align}
where $R(\epsilon|y_0^{l-1})$ can be set to $0$, $\epsilon$  is an empty sequence, and $$\tau(S,v)=\begin{cases}
0,&\sgn(S)=(-1)^v\\
-|S|,&\text{otherwise}.
\end{cases}$$ It can be verified that 
\begin{equation}
\label{mEWIdentity}
\sum_{\beta = 0}^{2^{j}-1} \tau(S_{0}^{(0)}(y_\beta),c_\beta) = \sum_{\beta = 0}^{2^{j}-1} \tau(S_j^{(\beta)}(v_0^{\beta-1},y_0^{2^j-1}), v_\beta),
\end{equation}
where $c=v_0^{2^j-1}F_2^{\otimes j}$.


It was suggested in \cite{trifonov2014binary} to express  values $\W^{(i)}_1(u_0^{i}|y_0^{l-1} )$ via $W^{(j)}_{t}(v_0^{j}|y_0^{l-1})$ for some $j$. One can represent the kernel $K$ as $K = TF_{2}^{\otimes t}$, where $T$ is an $l \times l$ matrix. Let $v_0^{l-1}=u_0^{l-1}T$. Then, $c_{0}^{l-1} = v_0^{l-1} F_{2}^{\otimes t} =u_{0}^{l-1}K$, so that 
$u_0^{l-1} = v_0^{l-1}T^{-1}.$

Observe, that it is possible to reconstruct $u_0^i$ from $v_0^{\tau_i}$, where $\tau_i$ is the position of the last non-zero symbol in the $i$-th row of $T^{-1}$. Recall that successive cancellation decoding of polar codes with arbitrary kernel requires one to compute values $\W^{(i)}_1(u_0^{i}|y_0^{l-1} ), u_i\in \F_2$. However, fixing the values $u_0^{i-1}$ may impose constraints on $v_j, j > \tau_i$, which must be taken into account while computing these probabilities.

Indeed, vectors $u_0^{l-1}$ and $v_0^{l-1}$ satisfy the  equation
$$\Theta'(u_{l-1}\quad \dots \quad u_1\quad u_0\quad v_0\quad v_1 \quad \dots \quad v_{l-1})^{T}=0,$$
where $\Theta'=(\mathbb S\quad I)$, and $l\times l$ matrix $\mathbb S$ is obtained by transposing $T$ and reversing the order of columns in the obtained matrix. By applying elementary row operations, matrix $\Theta'$
can be transformed into a minimum-span form $\Theta$, such that the  first and last non-zero elements of the $i$-th row are located in  columns $i$ and  $z_i$, respectively, where all $z_i$ are distinct. This enables one to obtain symbols of vector $u$ as 
\begin{align}
\label{fTransformMinSpan}
u_{i}=\sum_{s=0}^{i-1} u_s\Theta_{l-1-i,l-1-s} +\sum_{t=0}^{j_i}v_t \Theta_{l-1-i, l+t},
\end{align}
where $j_i=z_{l-1-i}-l$. 
Let $h_i = \underset{0 \leq i' \leq i}{\max} j_{i'}$. It can be seen that\footnote{The method given in \cite{buzaglo2017efficient} is a special case of this approach. }
\begin{align}
\W^{(j)}_1(u_0^{j}|y_0^{l-1}) &= 
\sum_{v_0^{h_{j}}\in\mathcal Z_j} W^{(h_j)}_{t}(v_0^{h_{j}}|y_0^{l-1})
\nonumber \\  
&=\sum_{v_0^{h_{j}}\in\mathcal Z_j}\sum_{v_{h_j+1}^{l-1}} W_t^{(l-1)}(v_0^{l-1}|y_0^{l-1}),
\label{mKernelLook}
\end{align}
where $\mathcal Z_j$ is the set of vectors $v_0^{h_j}$, such that \eqref{fTransformMinSpan} holds for $i\in[j]$. 
Similarly we can rewrite the above expression for the case of the approximate probabilities
\begin{align}
\widetilde \W_1^{(j)}(u_0^j|y_0^{l-1}) &= \max_{v_0^{h_{j}}\in\mathcal Z_j}  \widetilde W^{(h_j)}_{t}(v_0^{h_{j}}|y_0^{l-1}) \nonumber \\
&= \max_{v_0^{h_{j}}\in\mathcal Z_j} \max_{v_{h_j+1}^{l-1}} W^{(l-1)}_{t}(v_0^{l-1}|y_0^{l-1}).
\end{align} 
 Let 
$
\mathcal Z_{i,b} = \set{v_0^{h_i}|v_0^{h_i}\in \mathcal Z_i, \text{where }u_i=b}
$. Hence, one obtains
\begin{align}
\bS_{1, i} = \max_{v_0^{h_{i}}\in\mathcal Z_{i,0}} R(v_0^{h_i}|y_0^{l-1}) -
\max_{v_0^{h_{i}}\in\mathcal Z_{i,1}} R(v_0^{h_i}|y_0^{l-1}).
\label{mKernLLR}
\end{align}
Observe that computing these values requires considering  multiple vectors $v_0^{h_i}$ of input symbols of the Arikan transform $F_2^{\otimes t}$. Let $\mathcal D_i = \{0,\dots,h_i\}\backslash \{j_0,\dots,j_i\}$ be a \textit{decoding window}, i.e. the set of indices of Arikan input symbols $v_0^{h_i}$, which are not determined by symbols $u_0^{i-1}$. The number of such vectors, which determines the decoding complexity,  is  $ 2^{|\mathcal D_i|}$. In general, one has $|\mathcal D_i|=O(l)$ for an arbitrary kernel. 

\section{Efficient processing of $16\times 16$ kernels}
\label{sKern16Proc}
 To minimize complexity of proposed approach $\eqref{mKernLLR}$ one needs to find kernels with small decoding windows while preserving required polarization rate ($>0.5$ in our case) and scaling exponent.
By computer search, based on heuristic algorithm presented in \cite{fazeli2014scaling}, we found a $16 \times 16$ kernel\\
\scalebox{0.8}{\parbox{0.6\textwidth}{
$$\setlength{\arraycolsep}{2pt}K_1=
\left(\begin{array}{cccccccccccccccc}
1&0&0&0&0&0&0&0&0&0&0&0&0&0&0&0\\
1&1&0&0&0&0&0&0&0&0&0&0&0&0&0&0\\
1&0&1&0&0&0&0&0&0&0&0&0&0&0&0&0\\
1&0&0&0&1&0&0&0&0&0&0&0&0&0&0&0\\
1&0&0&0&0&0&0&0&1&0&0&0&0&0&0&0\\
1&1&0&0&0&0&0&0&1&1&0&0&0&0&0&0\\
1&1&0&0&1&1&0&0&0&0&0&0&0&0&0&0\\
1&1&1&1&0&0&0&0&0&0&0&0&0&0&0&0\\
1&0&0&0&1&0&0&0&1&0&0&0&1&0&0&0\\
1&0&1&0&0&1&1&0&1&1&0&0&0&0&0&0\\
0&1&1&0&1&1&0&0&1&0&1&0&0&0&0&0\\
1&1&1&1&1&1&1&1&0&0&0&0&0&0&0&0\\
1&1&1&1&0&0&0&0&1&1&1&1&0&0&0&0\\
1&1&0&0&1&1&0&0&1&1&0&0&1&1&0&0\\
1&0&1&0&1&0&1&0&1&0&1&0&1&0&1&0\\
1&1&1&1&1&1&1&1&1&1&1&1&1&1&1&1\\
\end{array}
\right)
$$}}
with  BEC scaling exponent  $\mu(K_1) = 3.346$ \cite{fazeli2014scaling}. Furthermore, to minimize the size of decoding windows, we derived another kernel
$K_2 = P_{\sigma}K_1$, were $P_{\sigma}$ is a permutation matrix corresponding to permutation $\sigma = [0,1,2,7,3,4,5,6,9,10,11,12,8,13,14,15]$, with scaling exponent  $\mu(K_2) = 3.45$. Both kernels have polarization rate  $0.51828$.

\begin{table}
\caption{Input symbols $u_\phi$ for kernels $K_1, K_2$ as functions of  input symbols $v$ for $F_2^{\otimes 4}$}
\label{tDecWin}
\centering
\scalebox{0.82}{
\begin{tabular}{|l|p{0.09\textwidth}|c|c||p{0.09\textwidth}|c|c|}
\hline
\multirow{2}{*}{$\phi$}&\multicolumn{3}{c||}{$K_1$}&\multicolumn{3}{c|}{$K_2$}\\\cline{2-7}
   & $u_\phi $         & $\mathcal D_\phi$ & Cost& $u_\phi$      & $\mathcal D_\phi$ & Cost\\ \hline
0  & $v_0$              & $\set{}$ &15& $v_0$                      &$\set{}$&15\\ \hline
1  & $v_1$              & $\set{}$ &1& $v_1$                      &$\set{}$  &1\\ \hline
2  & $v_2$              & $\set{}$ &3& $v_2$                      & $\set{}$ &3\\ \hline
3  & $v_4$              & $\set{3}$ &21& $v_3$                     &$\set{}$  &1\\ \hline
4  & $v_8$              & $\set{3,5,6,7}$ &127& $v_4$               &$\set{}$  &7\\ \hline
5  & $v_6\oplus v_9$      & $\set{3,5,6,7}$ &48& $v_8$               & $\set{5,6,7}$ &67\\ \hline
6  & $v_5\oplus v_6 \oplus v_{10} $   & $\set{3,5,6,7}$ &95& $v_6\oplus v_9$       & $\set{5,6,7}$ &24\\ \hline
7  & $v_3$              & $\set{5,6,7}$ &1& $v_5\oplus v_6\oplus v_{10}$      & $\set{5,6,7}$ &47\\ \hline
8  & $v_{12}$           & $\set{5,6,7,11}$ &127& $v_{6}$              & $\set{5,7}$ &1\\ \hline
9  & $v_{6}$           & $\set{5,7,11}$ &1& $v_{10}$             & $\set{7}$ &1\\ \hline
10 & $v_{10}$              & $\set{7,11}$ &1& $v_7$                  & $\set{}$ &1\\ \hline
11 & $v_7$              & $\set{11}$ &1& $v_{11}$                 & $\set{}$ &1\\ \hline
12 & $v_{11}$           & $\set{}$ &1& $v_{12}$                   & $\set{}$ &7\\ \hline
13 & $v_{13}$           & $\set{}$ &1& $v_{13}$                   & $\set{}$ &1\\ \hline
14 & $v_{14}$           & $\set{}$ &3& $v_{14}$                   & $\set{}$ &3\\ \hline
15 & $v_{15}$           & $\set{}$ &1& $v_{15}$                   & $\set{}$ &1\\ \hline
\end{tabular}}
\end{table}

Table \ref{tDecWin} presents the right hand side of expression \eqref{fTransformMinSpan} for each $ i \in [16]$, as well as the corresponding decoding windows $\mathcal D_i$, for both kernels. It can be seen that the maximal  decoding windows size for  $K_1$ and $K_2$ is  $4$ and $3$, respectively. Note that by applying the row permutation to $K_1$, we have reduced decoding windows, but increased scaling exponent. Below we present efficient methods for computing  some input symbol LLRs for these   kernels.

\subsection{Processing of kernel $K_1$ with $\mu = 3.346$}
\label{sDecK} 
It can be seen that for $\phi \in \set{0,1,2,13,14,15}$ one has  $\bS_{1,\phi} =S_4^{(\phi)}(v_0^{\phi},y_0^{15})$, i.e. LLR for $F_2^{\otimes 4}$.
\subsubsection{phase 3}
In case of $\phi = 3$ expressions \eqref{mKernLLR} and \eqref{fTransformMinSpan} imply that the decoding window $\mathcal D_3 = \set{3}$ and LLR for $u_3$ is given by
$$
\bS_{1,3} = \max_{v_3} R( \hat v_0  \hat v_1  \hat v_2 v_3 0|y_0^{15}) -
\max_{v_3} R( \hat v_0  \hat v_1 \hat v_2 v_3 1|y_0^{15}), 
$$
where $ \hat v_ i= \hat u_i,i\in[3]$, are already estimated symbols.

To obtain LLR $\mathbf S_{1,3}$ one should compute:
\begin{itemize}
\item $S_4^{(3)}(v_0^{2},y_0^{15})$ with 1 operation,
\item $R(\hat v_0 \hat v_1 \hat v_2 v_3|y_0^{15})$ for $v_3 \in [2]$. Observe that this can be done with 1 summation, since $R(\hat v_0 \hat v_1 \hat v_2 v_3|y_0^{15}) = R(\hat v_0 \hat v_1 \hat v_2|y_0^{15}) + \tau(S_4^{(3)}(v_0^{2},y_0^{15}),v_3)$ and there is $v_3$ such as $\tau(S_4^{(3)}(v_0^{2},y_0^{15}),v_3) = 0$,
\item $S_4^{(4)}(v_0^{3},y_0^{15})$ for $v_3 \in [2]$ with $7 * 2$ operations,
\item $R( \hat v_0  \hat v_1  \hat v_2 v_3 v_4|y_0^{15})$ for $v_3,v_4 \in [2]$ with 2 operations,
\item $\max_{v_3} R( \hat v_0  \hat v_1  \hat v_2 v_3 v_4|y_0^{15})$ for $v_4 \in [2]$ with 2 operations,
 \item $\max_{v_3} R( \hat v_0  \hat v_1  \hat v_2 v_3 0|y_0^{15}) -
\max_{v_3} R( \hat v_0  \hat v_1 \hat v_2 v_3 1|y_0^{15})$ with 1 operation.
\end{itemize}
Total number of operations is given by 21.
\subsubsection{phase 4}
The decoding window is given by $\mathcal D_4 = \set{3,5,6,7}$ and 
$$
\bS_{1, 4} = \max_{v_0^{8}\in\mathcal Z_{4,0}} R(v_0^{8}|y_0^{15}) -
\max_{v_0^{8}\in\mathcal Z_{4,1}} R(v_0^{8}|y_0^{l-1}),
$$
where $\mathcal Z_{4,b}$ is given by the set of vectors $[\hat v_0 \hat v_1 \hat v_2 v_3 \hat v_4 v_5 v_6 v_7 b]$, $v_3,v_5,v_6,v_7 \in [2]$.
   
Instead of exhaustive enumeration of vectors $v_0^{h_{j}}$ in \eqref{mKernLLR}, we propose to exploit the structure of $K_1$  to identify some common subexpressions (CSE) in formulas for $R(v_0^{h_i}|y_0^{l-1})$ and $S_\lambda^{(i)} = S_\lambda^{(i)}(v_0^{i-1},y_0^{l-1})$, which can be computed once and used multiple times. In some cases computing these subexpressions reduces to decoding of well-known codes, which can be implemented with appropriate fast algorithms. Furthermore, we observe that the set of possible values of these subexpressions is less than the number of different $v_0^{h_{j}}$ to be considered. This results in further complexity reduction. More accurate and detailed description of CSE can be found in \cite{trofimiuk2019reduced}. To demonstrate this approach, we consider computing the LLR for $u_4$ of $K_1$.

This requires considering 16 vectors $v_0^7$ satisfying \eqref{fTransformMinSpan}.  According to \eqref{mScoreMinSum}, one obtains 
$
R(v_0^8|y_0^{15})=R(v_0^{7}|y_0^{15})+\tau\left(S_4^{(8)}(v_0^{7},y_0^{15}),v_8\right).
\label{v8K1}
$
Observe that $\set{v_0^7F_2^{\otimes 3}|v_0^7\in \bar {\mathcal Z}_4}$ is a coset of Reed-Muller code $RM(1,3)$, where  $\bar{\mathcal Z}_4$ is the set of vectors $v_0^{7}$, so that \eqref{fTransformMinSpan} holds for $i\in[4]$.  Furthermore,  $$R(v_0^{7}|y_0^{15})
=\frac{1}{2}\left(\sum_{i=0}^7(-1)^{c_i}s_i-\sum_{i=0}^7|s_i|\right),$$
where $s_i = S_{1}^{(0)}(\epsilon,(y_{i},y_{i+8})), i \in [8]$, $c_0^7= v_0^7F_2^{\otimes 3}$.  Assume for the sake of simplicity that $v_j=0,j\in\set{0,1,2,4}$. Then the first term in this expression can be obtained for each $v_0^7\in \bar {\mathcal Z}_4$ via the fast Hadamard transform (FHT) \cite{beery1986optimal} of vector $s$, and the second one does not need to be computed, since it cancels in \eqref{mKernLLR}.

 It remains to compute $S_4^{(8)}(v_0^{7},y_0^{15}),v_0^{7} \in \bar{\mathcal Z_4}$ and $|\bar{ \mathcal Z_4}| = 16$. In a straightforward implementation, one would recursively apply formulas \eqref{mMinSum1} and \eqref{mMinSum2} to compute $S_4^{(8)}$ for 16 vectors $v_0^{7}$. It appears that there are some CSE arising in this computation.

 At first, one needs to compute $S_{1}^{(1)}(c_i,y_i,y_{i+8}),i\in[8]$,  $c_0^{7} = v_0^7 F_2^{\otimes 3}$. Since $c_i \in \set{0,1}$,  the values $S_{1}^{(1)}(j,y_i,y_{i+8}),j\in \set{0,1},i\in[8]$
constitute the first set of CSE. We store them in the array $L$ 
  $$L[4i+j]=S_{1}^{(1)}(j\bmod 2 ,y_{i+\bar j},y_{i+8+\bar j}),$$
 where $i,j \in [4]$, $\bar j = 4\lfloor j/2\rfloor$. Computing these values  requires $16$ summations only, instead of $16\cdot 8 = 128$ summations in a straightforward implementation.

The next step is to compute the values $S_2^{(2)}((c_i,c_{i+4}),(y_i,y_{i+4},y_{i+8},y_{i+12})), i\in[4]$ which are equal to  $Q(S_{1}^{(1)}(c_i,y_i,y_{i+8}),S_{1}^{(1)}(c_{i+4},y_{i+4},y_{i+12}))$. Since  $(c_i,c_{i+4}) \in \F_2^2$, $S_2^{(2)}$ gives us the second set of CSE. One can use values stored in $L$ to compute $S_2^{(2)}$ as 
$$X[i][j] = Q(L[4i+j/2], L[4i +(j \text{ mod }2) + 2])], i,j \in [4].$$

Observe that for any $c_0^7\in RM(1,3)$  one has $c_i^{i+4}\in RM(1,2), i\in\set{0,4}$. That is, one needs to consider only vectors $c_i^{i+4}$ of even weight while computing $S_3^{(4)}$. These values can be calculated as 
$$Y[i][j+4k] = Q(X[i][j \oplus 3k],X[i+2][j]),i,k \in [2], j\in[4].$$

Finally, the values $S_4^{(8)}(v_0^{7},y_0^{15})$ can be obtained as $$Z[i+8j] = Q(Y[0][i \oplus 3j],Y[1][i]), i \in[8], j\in [2]. $$ Each element of $Z$ corresponds to some $c_0^7\in RM(1,3)$.  Finally, these values are used in  \eqref{mKernLLR} together with $R(v_0^{7}|y_0^{15})$ to calculate $\mathbf S_{1,4}$.

Let us compute the number of operations required to process phase $4$. One need to compute
\begin{itemize}
\item $R(v_0^7|y_0^{15})$ for $v_0^7 \in \bar {\mathcal Z}_4$ via FHT with 24 operations,
\item all different $S_1^{(1)}$ arising in CSE (array $L$)\ with 16 operations,
\item all $S_2^{(2)}$ in CSE (array $X$) with 16 operations,
\item all $S_3^{(4)}$ in CSE (array $Y$) with 16 operations,
\item all $S_4^{(8)}$ in CSE (array $Z$) with 16 operations,
\item $R(v_0^8|y_0^{15})$ for $v_0^8 \in {\mathcal Z}_4$ with 16 operations,
\item $\max_{v_0^{8}\in\mathcal Z_{4,b}} R(v_0^{8}|y_0^{15})$, $b \in [2]$, with $15 * 2$ operations,
\item $\mathbf S_{1,4}$ with 1 operation. 
\end{itemize}

The overall complexity is given by 135 operation. 

We also employ one observation to reduce the complexity of computing $\max_{v_0^{8}\in\mathcal Z_{4,b}} R(v_0^{8}|y_0^{15})$. Let $s$ be a FHT\ of the vector $s$, where $s_i = S_{1}^{(0)}(\epsilon,(y_{i},y_{i+8})), i \in [8]$. Observe that we can compute $\arg\max_{i \in [8]} |s_i|$ with 7 operations and obtain $v_3,v_5,v_6,v_7$   which gives us $\max R(v_0^7|y_0^{15})$. Recall that $\tau(S,c)$ function is zero for one of $b \in [2]$, therefore, there is a value of $\hat b, b \in [2]$ such as $\max_{v_0^{8}\in\mathcal Z_{4,\hat b}} R(v_0^{8}|y_0^{15}) = \max R(v_0^7|y_0^{15})$. It implies that we remain need to compute $\max_{v_0^{8}\in\mathcal Z_{4,1 \oplus \hat b}} R(v_0^{8}|y_0^{15})$. With this modification we have the complexity given by 127 operations.

\subsubsection{phase 5}
The decoding window $\mathcal D_5$ remains the same as in the previous phase. 
According to expressions \eqref{mMinSum2} and \eqref{mScoreMinSum} to obtain $\mathbf S_{1,5}$ one should compute:
\begin{itemize}
\item all $S_4^{(9)}$ with 16 operations,
\item $R(v_0^9|y_0^{15})$ for $v_0^9 \in {\mathcal Z}_5$ with 16 operations,
\item $\max_{v_0^{9}\in\mathcal Z_{5,b}} R(v_0^{9}|y_0^{15}), b \in [2],$ with 15 operations.  Similarly to phase 4, there is a value of $\hat b \in [2]$ such as $\max_{v_0^{9}\in\mathcal Z_{5,\hat b}} R(v_0^{9}|y_0^{15}) = \max_{v_0^{8}\in\mathcal Z_{4,v_4}} R(v_0^{8}|y_0^{15})$,
\item $\mathbf S_{1,5}$ with 1 operation.
\end{itemize}  

Total complexity is given by 48 operations.
\subsubsection{phase 6}
The decoding window $\mathcal D_6 = \set{3,5,6,7}$. 

At this phase according to expressions \eqref{mMinSum1}-\eqref{mMinSum2} one should compute:
\begin{itemize}
\item 32 values of $S_3^{(5)}$ in CSE with 32 operations,
 \item 16 LLRs $S_4^{(10)}$ with 16 operations,
    \item $R(v_0^{10}|y_0^{15})$ for $v_0^{10} \in {\mathcal Z}_6$ with 16 operations,
    \item $\max_{v_0^{10}\in\mathcal Z_{6,b}} R(v_0^{10}|y_0^{15}), b \in [2],$ with 30 operations,
\item $\mathbf S_{1,5}$ with 1 operation.
\end{itemize}
Total complexity is given by 95 operations.
\subsubsection{phase 7}
At this phase the decoding window is reduced and given by $\mathcal D_7 = \set{5,6,7}$. Moreover, the value $h_7 = h_6 = v_{10}$, which means that the values $R(v_0^{10}|y_0^{15})$ remains the same. We propose to use the following method: at phase 6 one should compute $\max_{\set{v_0^{10}|v_0^{10}\in\mathcal Z_{7,b},u_6 = \bar b}} R(v_0^{10}|y_0^{15})$ for $b \in [2], \bar b \in [2]$ and obtain 
$$\max_{v_0^{10}\in\mathcal Z_{6,b}} R(v_0^{10}|y_0^{15}) = \max_{\bar b \in [2]} \max_{\set{v_0^{10}|v_0^{10}\in\mathcal Z_{7,b},u_6 = \bar b}} R(v_0^{10}|y_0^{15}).$$

Once the value $u_6$ is determined, one can obtain $\mathbf S_{1,7}$ with one operation directly from already computed values $\max_{v_0^{10}\in\mathcal Z_{7,b}} R(v_0^{10}|y_0^{15})$.
\subsubsection{phase 8}
At this phase the decoding window is increased and given by $\mathcal D_8 = \set{5,6,7,11}$ and $h_8 = 12$. To obtain kernel input symbol LLR one should compute:

\begin{itemize}
    \item 8 LLRs $S_4^{(11)}$ with 8 operations,
    \item 16 path scores $R(v_0^{11}|y_0^{15})$ with 8 operations,
    \item 16 values $S_2^{(3)}$ in CSE with 16 operations,
    \item 32 values $S_3^{(6)}$ in CSE with 32 operations,
    \item 16 values $S_4^{(12)}$ with 16 operations,
    \item 32 path scores $R(v_0^{12}|y_0^{15})$ with 16 operations,
    \item $\max_{v_0^{12}\in\mathcal Z_{12,b}} R(v_0^{12}|y_0^{15}), b\ in [2]$ with $15 * 2$ operations,
\item $\mathbf S_{1,12}$ with one operation.
\end{itemize}
Total complexity is given by 127 operations. 

One can recursively apply approach described for phase 7, namely, construct tree of maximums of $R(v_0^{12}|y_0^{(15)})$, and obtain $\mathbf S_{1,\phi}, 8<\phi<13$ with one operation.

\subsection{Processing of kernel $K_2$ with $\mu = 3.45$}
Below we briefly present a complete  processing algorithm for kernel  $K_2$. It provides much better performance-complexity tradeoff compared to $K_1$. The algorithm uses the same CSE elimination techniques  as  described in section \ref{sDecK}. After the pre-computation steps for $\phi\in\set{5,6,7},$  the LLR is obtained via \eqref{mKernLLR}.
 

\begin{itemize}
\item For $\phi \in \set{0,1,2,3, 4, 11,12, 13,14,15}$, compute $\bS_{1,\phi}$ as LLRs  $S_4^{(\phi)}(v_0^{\phi-1},y_0^{15})$ for the Arikan transform $F_2^{\otimes 4}$.
\item For $\phi = 5$:
\begin{enumerate}
\item Since the set of vectors 
$c_0^3 = [\hat v_4 v_5 v_6 v_7]F_2^{\otimes 2}$
fixed $v_i,i\in\set{0,1,2,3,4},$ a coset of  $RM(1,2)$, one can obtain $8$ values of  $R(v_4^7|y_0^{15})$ (recall that 
$R(v_0^7|y_0^{15}) = R(\hat v_0^3|y_0^{15}) + R(v_4^7|y_0^{15})$) 
from the FHT of the vector $s_i = S_{2}^{(1)}(\bar c_i, (y_i,y_4,y_{i+8},y_{i+12})), i \in [4]$, $\bar c = \hat v_0^3 F_2^{\otimes 2}$.
\item Compute\\  $L[i+8j]=S_{1}^{(1)}(j\bmod 2 ,y_{i},y_{i+8}), i \in [8], j \in [2].$
\item Since $\set{(c_i,c_{i+4})}=\set{(0,0),(1,1)}$, compute all possible  $S_2^{(2)}$ values as 

$X[i][j] = Q(L[i+8j],L[i+4+8j]), i \in [4], j \in [2]. $
\item Since $\set{(c_i,c_{i+4},c_{i+2},c_{i+6})}$ is a code generated by $\begin{pmatrix}1&1&0&0\\1&1&1&1\end{pmatrix}$, compute all possible $S_3^{(4)}$ values as 

$Y[i][j] = Q(X[i][j/2],X[i+2][j\mod2]), i \in [2], j \in[4].$ 
\item For every $c_0^7=v_0^7F_2$ compute $S_4^{(8)}(v_0^7,y_0^{15})$ as 

$Z[i+4j] = Q(Y[0][i\oplus3j],Y[1][i]), i \in [4], j \in [2].$
\end{enumerate}
\item For $\phi = 6$, compute $S_4^{(9)}(v_0^8,y_0^{15})$  as 

$Z[i+4j] = P(Y[0][i\oplus3j],Y[1][i],v_{8}), i \in [4], j \in [2]$.
\item For $\phi = 7$:
\begin{enumerate}
\item Compute $S_3^{(5)}$ as 

$\bar Y[0][j+4k] = P(X[0][j/2],X[2][j\text{ mod }2],\bar c_{0,k})$,
$\bar Y[1][j+4k] = P(X[3][\bar j_k/2],X[1][\bar j_k \text{ mod }2],\bar c_{1,k})$

 and $\bar c_{0, k} = v_8\oplus u_6 \oplus k$, $\bar c_{1,k} = u_6 \oplus k$, 
$\bar j_k = j \oplus 3$, $i \in [2], j \in[4], k \in[2]$.
\item Obtain $S_4^{(10)}$ as 
$Z[i] = Q(\bar Y[0][i], \bar Y[1][i]), i \in [8]$.
\end{enumerate}
\item For $\phi \in \set{8,9,10}$:
Use $16$ already computed values of $R(v_0^{10}|y_0^{15})$ to obtain $\bS_{1,\phi}$.
\end{itemize}
\begin{remark}
Let us comment the case of $\phi = 7$. In conventional Arikan SC for $F_2^{\otimes 4}$, after symbol $v_9$ is estimated, the LLRs $S_3^{(5)}$ for $v_{10}$ are obtained by applying $P$ function to   $S_2^{(2)}$ and values $(v_8\oplus v_9, v_8)$. In the case of $K_2$, we do not have a fixed value for $v_9$. Instead, we have a constraint $u_6 = v_6+v_9 $. Therefore, for each vector $v_0^9 \in \bar{\mathcal Z_7}$ the value of $v_9$ is changed according to $v_6$. This property is taken into account in the expressions for computing of  $\bar Y[i][j]$. 
\end{remark}

For processing of $K_2$ we also used trick with simplified computation of path score maximums similarly to phase $5$ of $K_1$.

The cost, in terms of the total number of summations and comparisons, of computing $\bS_{1, \phi}$ using the proposed algorithm is shown in Table \ref{tDecWin}. The overall processing complexity is $447$ and $181$ operations for kernels $K_1$ and $K_2$ respectively, while the trellis-based algorithm \cite{griesser2002aposteriori} requires $7557$ and $9693$ operations, respectively.

The above described techniques can be also used to implement an SCL\ decoder for polar codes with the considered kernels, using a straightforward generalization of the algorithm and data structures presented in \cite{tal2015list}.
\section{Numeric results}
\begin{figure}
\includegraphics[width=0.5\textwidth]{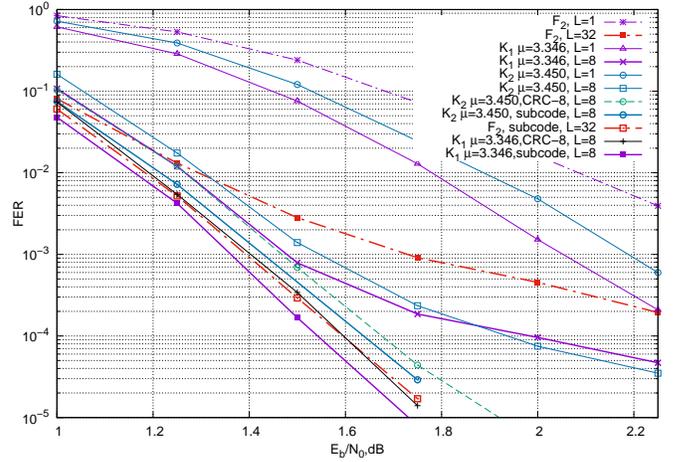}
\caption{Performance of $(4096, 2048)$ polar  codes}
\label{f4096_2048}
\end{figure}
We constructed $(4096,2048)$ polar codes with the considered kernels, and investigated their performance for the case of AWGN\ channel with BPSK\ modulation. The sets of frozen symbols were obtained by Monte-Karlo simulations.

Figure \ref{f4096_2048} illustrates the  performance of plain polar codes, polar codes with CRC\footnote{CRC length was selected to minimize FER with $L=8$.} and polar subcodes  \cite{trifonov2017randomized}. It can be seen that the codes based on kernels $K_1$ and $K_2$ with improved polarization rate $E(K_1) = E(K_2) = 0.51828$ provide significant performance gain compared to polar codes with Arikan kernel.  Observe also that randomized polar subcodes provide better performance compared to polar codes with CRC. Moreover, polar subcodes with kernels $K_1, K_2$ under SCL with $L=8$ have almost the same performance as polar subcodes with Arikan kernel under SCL with $L = 32$.  Observe also that the codes based on  kernels with lower scaling exponent exhibit better performance.

\begin{figure}
\begin{subfigure}[b]{0.2415\textwidth}
\includegraphics[width=\linewidth]{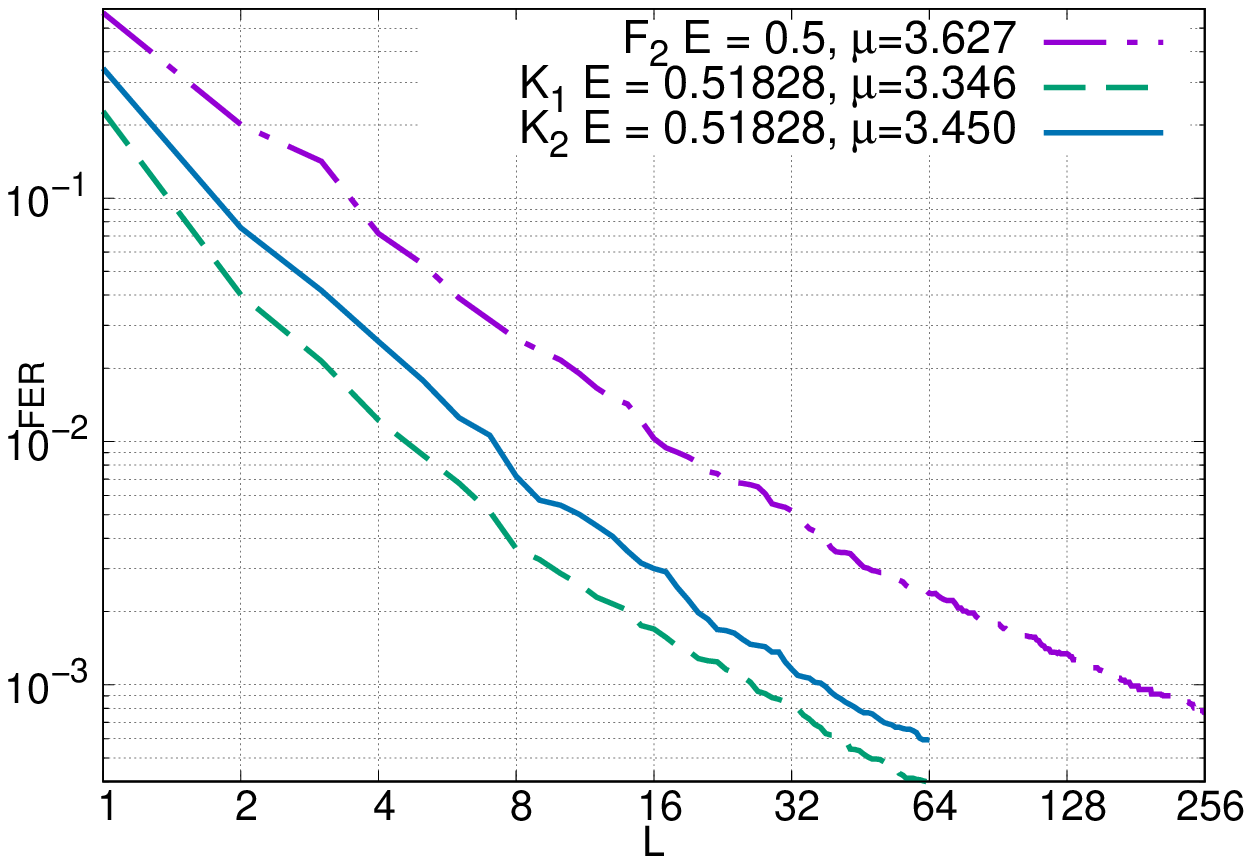}
\caption{Performance}
\label{fErrorList}
\end{subfigure}
\begin{subfigure}[b]{0.2415\textwidth}
\includegraphics[width=1\linewidth]{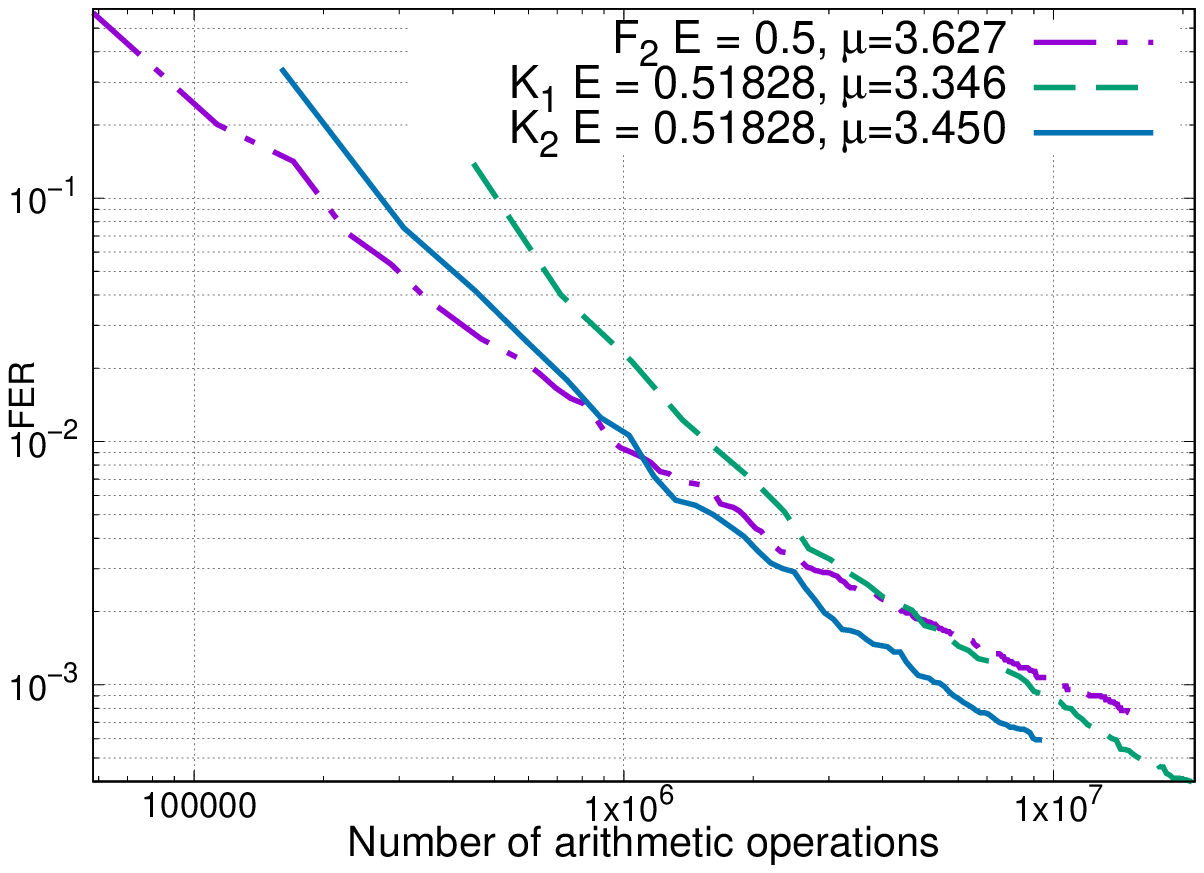}
\caption{Decoding complexity}
\label{fErrorCompl}
\end{subfigure}
\caption{SCL decoding of polar subcodes with different kernels}
\end{figure}

Figure \ref{fErrorList} presents simulation results for $(4096,2048)$ polar subcodes with different kernels under SCL with different $L$ at $E_b/N_0=1.25$ dB. It can be seen that the kernels with polarization rate $0.51828$ require significantly lower list size $L$ to achieve the same performance as the code with the Arikan kernel. Moreover, this gap grows with $L$. This is due to improved rate of polarization, which results in smaller number of unfrozen imperfectly polarized bit subchannels. The size of the list needed to correct possible errors in these subchannels grows exponentially with their number (at least for the genie-aided decoder considered in \cite{mondelli2015scaling}). On the other hand, lower scaling exponent gives better performance with the same list $L$, but the slope of the curve remains the same for both kernels $K_1,K_2$.

Figure \ref{fErrorCompl} presents the same results in terms of the actual decoding complexity. 
Recall that proposed kernel processing algorithm uses only summations and comparisons. The  SCL algorithm was implemented using the randomized order statistic algorithm  for selection of the paths to be killed at each phase, which has complexity $O(L)$. Observe that the polar subcode based on kernel $K_2$ can provide better performance with the same decoding complexity for FER $\leq 8\cdot 10^{-3}$. This is due to higher slope of the corresponding curve  in Figure \ref{fErrorList}, which eventually enables one to compensate relatively high complexity of the LLR computation algorithm presented in Section \ref{sKern16Proc}.
 
Unfortunately, $K_1$ kernel, which provides lower scaling exponent, has greater processing complexity than $K_2$, so that its curve intersects the one for  the Arikan kernel only at FER$=2\cdot 10^{-3}$.
\section{Conclusions}
In this paper efficient decoding algorithms for some $16 \times 16$ polarization kernels with polarization rate $0.51828$ were proposed. The algorithms compute kernel input symbols LLRs   via the ones for the Arikan kernel, and exploit the structure of the codes induced by the kernel to identify and re-use the values of some common subexpressions. It was shown that in the case of SCL decoding with sufficiently large list size, the proposed approach results in lower decoding complexity compared to the case of polar (sub)codes with Arikan kernel with the same performance.

Extension of the proposed approach to the case of other kernels remains an open problem.
\section*{Acknowledgment}
We thank Fariba Abbasi Aghdam Meinagh for many comments and stimulating discussions.

\bibliographystyle{ieeetran}

\begin{thebibliography}{10}
\providecommand{\url}[1]{#1}
\csname url@samestyle\endcsname
\providecommand{\newblock}{\relax}
\providecommand{\bibinfo}[2]{#2}
\providecommand{\BIBentrySTDinterwordspacing}{\spaceskip=0pt\relax}
\providecommand{\BIBentryALTinterwordstretchfactor}{4}
\providecommand{\BIBentryALTinterwordspacing}{\spaceskip=\fontdimen2\font plus
\BIBentryALTinterwordstretchfactor\fontdimen3\font minus
  \fontdimen4\font\relax}
\providecommand{\BIBforeignlanguage}[2]{{%
\expandafter\ifx\csname l@#1\endcsname\relax
\typeout{** WARNING: IEEEtran.bst: No hyphenation pattern has been}%
\typeout{** loaded for the language `#1'. Using the pattern for}%
\typeout{** the default language instead.}%
\else
\language=\csname l@#1\endcsname
\fi
#2}}
\providecommand{\BIBdecl}{\relax}
\BIBdecl

\bibitem{arikan2009channel}
E.~Arikan, ``Channel polarization: A method for constructing capacity-achieving
  codes for symmetric binary-input memoryless channels,'' \emph{IEEE
  Transactions on Information Theory}, vol.~55, no.~7, pp. 3051--3073, July
  2009.

\bibitem{tal2015list}
I.~Tal and A.~Vardy, ``List decoding of polar codes,'' \emph{IEEE Transactions
  On Information Theory}, vol.~61, no.~5, pp. 2213--2226, May 2015.

\bibitem{trifonov2016polar}
P.~Trifonov and V.~Miloslavskaya, ``Polar subcodes,'' \emph{IEEE Journal on
  Selected Areas in Communications}, vol.~34, no.~2, pp. 254--266, February
  2016.

\bibitem{trifonov2017randomized}
P.~Trifonov and G.~Trofimiuk, ``A randomized construction of polar subcodes,''
  in \emph{Proceedings of IEEE International Symposium on Information
  Theory}.\hskip 1em plus 0.5em minus 0.4em\relax Aachen, Germany: IEEE, 2017,
  pp. 1863--1867.

\bibitem{wang2016paritycheckconcatenated}
T.~Wang, D.~Qu, and T.~Jiang, ``Parity-check-concatenated polar codes,''
  \emph{IEEE Communications Letters}, vol.~20, no.~12, December 2016.

\bibitem{korada2010polar}
S.~B. Korada, E.~Sasoglu, and R.~Urbanke, ``Polar codes: Characterization of
  exponent, bounds, and constructions,'' \emph{IEEE Transactions on Information
  Theory}, vol.~56, no.~12, pp. 6253--6264, December 2010.

\bibitem{fazeli2018binary}
A.~Fazeli, S.~H. Hassani, M.~Mondelli, and A.~Vardy, ``Binary linear codes with
  optimal scaling: Polar codes with large kernels,'' in \emph{Proceedings of
  IEEE Information Theory Workshop}, 2018.

\bibitem{fazeli2014scaling}
A.~Fazeli and A.~Vardy, ``On the scaling exponent of binary polarization
  kernels,'' in \emph{Proceedings of 52nd Annual Allerton Conference on
  Communication, Control and Computing}, 2014, pp. 797 -- 804.

\bibitem{presman2015binary}
N.~Presman, O.~Shapira, S.~Litsyn, T.~Etzion, and A.~Vardy, ``Binary
  polarization kernels from code decompositions,'' \emph{IEEE Transactions On
  Information Theory}, vol.~61, no.~5, May 2015.

\bibitem{buzaglo2017efficient}
S.~Buzaglo, A.~Fazeli, P.~H. Siegel, V.~Taranalli, and A.~Vardy, ``On efficient
  decoding of polar codes with large kernels,'' in \emph{Proceedings of IEEE
  Wireless Communications and Networking Conference Workshops (WCNCW)}, March
  2017, pp. 1--6.

\bibitem{miloslavskaya2014sequentialBCH}
V.~Miloslavskaya and P.~Trifonov, ``Sequential decoding of polar codes with
  arbitrary binary kernel,'' in \emph{Proceedings of IEEE Information Theory
  Workshop}.\hskip 1em plus 0.5em minus 0.4em\relax Hobart, Australia: IEEE,
  2014, pp. 377--381.

\bibitem{griesser2002aposteriori}
H.~Griesser and V.~R. Sidorenko, ``A posteriory probability decoding of
  nonsystematically encoded block codes,'' \emph{Problems of Information
  Transmission}, vol.~38, no.~3, 2002.

\bibitem{miloslavskaya2014sequential}
V.~Miloslavskaya and P.~Trifonov, ``Sequential decoding of polar codes,''
  \emph{IEEE Communications Letters}, vol.~18, no.~7, pp. 1127--1130, 2014.

\bibitem{trifonov2018score}
P.~Trifonov, ``A score function for sequential decoding of polar codes,'' in
  \emph{Proceedings of IEEE International Symposium on Information Theory},
  Vail, USA, 2018.

\bibitem{trifonov2014binary}
------, ``Binary successive cancellation decoding of polar codes with
  {Reed-Solomon} kernel,'' in \emph{Proceedings of IEEE International Symposium
  on Information Theory}.\hskip 1em plus 0.5em minus 0.4em\relax Honolulu, USA:
  IEEE, 2014, pp. 2972 -- 2976.

\bibitem{trofimiuk2019reduced}
G.~{Trofimiuk} and P.~{Trifonov}, ``Reduced complexity window processing of
  binary polarization kernels,'' in \emph{Proceedings of IEEE International
  Symposium on Information Theory}, Paris, France, July 2019.

\bibitem{beery1986optimal}
Y.~Beery and J.~Snyders, ``Optimal soft decision block decoders based on fast
  {Hadamard} transform,'' \emph{IEEE Transactions on Information Theory},
  vol.~32, no.~3, May 1986.

\bibitem{mondelli2015scaling}
M.~Mondelli, S.~H. Hassani, and R.~Urbanke, ``Scaling exponent of list decoders
  with applications to polar codes,'' \emph{IEEE Transactions On Information
  Theory}, vol.~61, no.~9, September 2015.

\end{thebibliography}

\end{document}